\documentstyle[11pt,newpasp,twoside]{article}
\def\edcomment#1{\iffalse\marginpar{\raggedright\sl#1\/}\else\relax\fi}
\reversemarginpar

\begin{document}
\title{X-ray Cavities and Cooling Flows}
\author{Paul E.~J.~Nulsen}
\affil{School of Engineering Physics, University of Wollongong,
Wollongong NSW 2522, Australia, and
Harvard-Smithsonian Center for Astrophysics, 60 Garden St, Cambridge,
MA 02138, USA}
\author{Brian R.~McNamara}
\affil{Department of Physics and Astronomy, Ohio University, Athens,
OH 45701, USA}
\author{Laurence P.~David}
\affil{Harvard-Smithsonian Center for Astrophysics, 60 Garden St,
Cambridge, MA 02138, USA}
\author{Michael W.~Wise}
\affil{Massachusetts Institute of Technology, Center for Space
Research, 70 Vassar St, Building 37, Cambridge, MA 02139, USA}

\begin{abstract}
Recent data have radically altered the X-ray perspective on cooling
flow clusters.  X-ray spectra show that very little of the hot
intracluster medium is cooler than about 1 keV, despite having short
cooling times.  In an increasing number of cooling flow clusters, the
lobes of a central radio source are found to have created cavities in
the hot gas.  Generally, the cavities are not overpressured relative
to the intracluster gas, but act as buoyant bubbles of radio emitting
plasma that drive circulation as they rise, mixing and heating the
intracluster gas.  All this points to the radio source, i.e. an active
galactic nucleus, as the heat source that prevents gas from cooling to
low temperatures.  However, heating due to bubbles alone seems to be
insufficient, so the energetics of cooling flows remain obscure.  We
briefly review the data and theory supporting this view and discuss
the energetics of cooling flows.
\end{abstract}

\section{Introduction}

The radiative cooling time of the hot intergalactic gas close to the
centres of about 70 percent of rich clusters of galaxies is
significantly shorter than the Hubble time.  These systems are known
as cooling flows (Fabian 1994).  Over the lifetime of a cooling flow
cluster, radiative losses have a significant impact on the gas unless
the radiated heat is replaced.  Cooling gas is compressed in order to
maintain hydrostatic equilibrium, causing inflow near to the cluster
centre and the deposition of large quantities of cool gas.  X-ray data
from {\it Chandra} and {\it XMM-Newton} confirm central cooling times
as short as $10^8$ -- $10^9$ y in many clusters (e.g. David et {al.}
2001), highlighting the significance of radiative losses.  However,
100 -- $1000\rm\ M_\odot\ y^{-1}$ of cold gas should be deposited by
cooling flows (e.g.~White, Jones \& Forman 1997) and very little
evidence is found of this gas.  Many forms of cool gas (e.g. Crawford
et {al.} 1999; Edge 2001) and recent star formation (e.g. Mittaz et
{al.}  2001) have been found at the centres of cooling flow clusters,
but the amounts fall well short of those expected.  Cooling flows also
occur in groups and isolated elliptical galaxies (Mathews \& Brighenti
2003).

X-ray spectra from the Reflection Grating Spectrograph (RGS) on {\it
XMM-Newton} show that there is very little gas cooler than about 1 keV
in cluster cooling flows.  If the late stages of cooling are isobaric
as expected, then the luminosity in a line is $L_{\rm line} = \dot M
(5k/(2\mu m_{\rm H}) \int_0^{T_{\rm max}} \Lambda_{\rm line}(T) /
\Lambda(T) \, dT$, where $\dot M$ is the deposition rate of cooled
gas, $\Lambda$ is the cooling function and $\Lambda_{\rm line}$ the
part of the cooling function due to the line.  This prediction is
quite robust for low temperature lines, but RGS data show that some
low energy lines are at least an order of magnitude weaker than
expected (e.g. Peterson et {al.}  2003).

\section{Radio Lobe Cavities in Cooling Flows}

Burns (1990) found that 70 percent of cD galaxies in cooling flow
clusters are radio loud, compared to 25 percent overall.  Observations
with {\it Chandra} reveal a growing list of clusters where
radio lobes at the cluster centre have created cavities in the hot
intracluster gas.  Some examples are Perseus (B\"ohringer et
{al.}  1993; Fabian et {al.}  2000), Hydra A (McNamara et {al.} 2000),
Abell 2052 (Blanton et {al.}  2001), RBS797 (Schindler et {el.} 2001),
MKW3S (Mazzotta et {al.}  2002) and Abell 4059 (Heinz et {al.} 2002).

Contrary to expectation, there is little evidence of shocks driven by
expanding radio lobes in most systems (but see Kraft et {al.} 2003;
Fabian et {al.}  2003).  Furthermore, the coolest X-ray emitting gas
surrounds the cavities in many systems (e.g.~Nulsen et {al.}  2002).
It would be very surprising to find the lowest entropy gas close to
the origin of a strong shock.  Lastly, the equipartition pressure of
the radio lobes is typically about one tenth of the surrounding gas
pressure.  All this argues that the radio lobes are at nearly the same
pressure as the surrounding gas.

The cool gas around the cavities is surprising.  Nulsen et {al.}
(2002) argued that this is not due to shock induced cooling or
magnetohydrodynamic shocks in Hydra A.  However, in a temperature map,
they found a plume of cool gas extending from the centre to beyond the
radio lobe cavities in Hydra A.  This suggests that repeated radio
outbursts have produced buoyant bubbles (cavities) that drive outflow
along the radio axis, lifting some low entropy gas from the cluster
centre.  Numerical simulations support this model (Br\"uggen et
{al.} 2002).

\section{Energetics}

The question of what prevents gas from cooling to low temperatures in
cooling flows remains a major issue.  The heat required to make up for
radiative losses from the region where the cooling time is shorter
than the age of a cluster is typically $10^{44}$ -- $10^{45}\rm\ erg\
s^{-1}$.  Also, a significant amount of gas at the cluster centre must
be maintained with coolings times of $ 10^8$ -- $10^9$ y.  This is
very difficult to achieve without a process that involves feedback.

Radio lobes are powered by an AGN, which is likely fuelled by cooled
and cooling gas.  If the mechanical energy input due to the cavities
is appreciable, this provides a feedback mechanism linking cooling and
AGN heating.  We consider the energy input by the bubbles, using
Hydra A for illustration (David et {al.}  2001).  The work
of creating a bubble in local pressure equilibrium is $p V \simeq 2.8
\times 10^{59}$ erg for the SW cavity of Hydra A.  The free energy of
a bubble is the sum of this and its thermal energy, giving the
enthalpy $\gamma pV /(\gamma-1)$, where $\gamma$ is the ratio of
specific heats.  We double this again to allow for the NE bubble.  If
the cavity is dominated by relativistic plasma, then $\gamma=4/3$
and the total free energy of the two cavities in Hydra A $\simeq
2.2\times10^{60}$ erg.  If all of this is thermalized within the
cooling flow region of Hydra A, then it can prevent gas from cooling
for $2.3\times 10^8$ y.

Churazov et {al.} (2002) argue that the enthalpy of a rising adiabatic
bubble decreases with pressure, and the loss goes into heating the
gas.  A bubble rises as the ICM falls in around it, converting
potential energy to kinetic energy, which is then dissipated in the
bubble's wake.  The potential energy released when a bubble of volume
$V$ rises a distance $\delta R$ is $\delta E = \rho V g \, \delta R$,
where $\rho$ is the density of the ICM and $g$ the acceleration due
to gravity.  From the equation of hydrostatic equilibrium, $\rho g = -
dp/dr$, so that $\delta E = - V dp/dr \, \delta R = - V \, \delta p$.
For an adiabatic bubble, $pV^\gamma =$ constant, and this result is
readily integrated, giving the energy dissipated over a finite length
of wake, $\Delta E = H_0 - H$, where the enthalpy depends on the
pressure through $H = H_0 (p/p_0)^{(\gamma-1)/\gamma}$.  For Hydra A,
with $\gamma = 4/3$, about half of the free energy is dissipated in
the cooling flow region.  If the bubble is non-adiabatic, more energy
is deposited in the core.

While Hydra A is a very powerful FRI radio source, its cavities are
not exceptional.  Indeed, the existence of ``ghost'' cavities ({e.g.}
McNamara et {al.} 2001) tells us that radio lifetime can be shorter
than bubble lifetime and we should not expect a strong correlation
between bubble energy and radio power.  There may also be other energy
inputs from an AGN, including direct injection of relativistic
particles, uncollimated outflows, or Compton heating.

It is generally difficult to heat from the centre of a cluster without
creating a well mixed, isentropic core (Fabian et {al.} 2001;
Brighenti \& Mathews 2003).  Slow heating drives steady convection.
Fast heating drives a shock, causing entropy inversion, then
convection and mixing.  Mixing is less thorough if energy is deposited
off centre, forming bubbles.  However, to prevent the bulk of the
lowest entropy gas from cooling to low temperatures, most gas must be
heated substantially at some stage and so take part in large-scale
convection, tending to disrupt observed abundance gradients ({e.g.}
David et {al.}  2001).

Zakamska \& Narayan (2003) have shown that thermal conduction can
balance radiative losses in some, but not all cooling flows.  However,
since it involves no feedback, maintaining cool gas by thermal
conduction requires very fine tuning ({e.g.} Bregman \& David 1988).
Also, thermal conductivity must be suppressed to explain observed
structures in several clusters.  It has been argued that these are
special cases, but this is harder to accept for the large scale
suppression found by Markevitch et {al.} (2003).

Even when suppressed by a factor of 10 or more, thermal conduction can
balance radiative losses in the outer parts of cluster cooling flows.
Thus, thermal conduction may augment AGN heating in clusters.  In that
case, AGN heating need only account for radiative losses from near to
the cluster centre.  Thermal conduction falls rapidly with
temperature, and so is less likely to be significant in groups and
isolated elliptical galaxies.  On the other hand, their energetics are
less demanding, since a single AGN outburst can disrupt the hot
interstellar medium of an isolated elliptical ({e.g.} Finoguenov \&
Jones 2001).

It has long been argued that major mergers can completely disrupt a
cooling flow (McGlynn \& Fabian 1984) and a variety of mechanisms have
been proposed to tap the energy of mergers to prevent gas from cooling
({e.g.} Motl et {al.} 2003).  However, in minor mergers and infall,
most energy is deposited in the outer parts of clusters.  Stable
stratification, and the huge density and pressure contrast from centre
to edge are obstacles to getting energy from the outer regions
deposited in the cluster core.  Furthermore, without feedback it is
very difficult for such a process to maintain short cooling times in
the cluster core.  As for thermal conduction, these effects may
augment AGN heating.

\section{Discussion and Conclusions}

There is good evidence of AGN heating in cooling flows, but it remains
unclear whether it is significant for cooling flows as a whole.  Since
AGN heating is linked to cooling by feedback, this process can
plausibly explain how gas can be maintained with short cooling times
in cooling flows.  AGN heating probably needs to be augmented in order
to account for global energetics of cooling flows.  Many details of
the heating process remain obscure.  In particular, it is unclear how
a cluster can be heated from its centre without producing a constant
entropy core and mixing out observed abundance gradients.

To end on a speculative note, AGN outbursts also occur in isolated
elliptical galaxies, where they can prevent the cooling of hot gas
more readily than in clusters.  If so, AGN feedback inhibits cooling,
hence star formation, in almost any system dominated by hot gas.  In
that case, the effect of AGN feeback is imprinted on the galaxy
luminosity function.

\acknowledgments  This work was partly supported by the Australia
Research Council and by NASA grant NAS8-01130.

\end{document}